\documentclass[prb,aps,groupedaddress,amsmath,twocolumn]{revtex4-1}
\usepackage[dvips]{graphicx}
\usepackage{dcolumn}
\usepackage{bm}
\usepackage{color}
\begin{document}

\title{Quantum Monte Carlo calculations of van der Waals interactions between aromatic benzene rings} 

\author{Sam Azadi}
\email{s.azadi@imperial.ac.uk}
\affiliation{Department of Physics, Imperial College London, SW7 2AZ London, United Kingdom}

\author{T.\ D.\ K\"{u}hne}
\affiliation{Department of Chemistry and Paderborn Center for Parallel
  Computing, University of Paderborn, Warburger Str.\ 100, D-33098 Paderborn,
  Germany} 
\affiliation{Center for Sustainable Systems Design and Institute for Lightweight Design with Hybrid Systems, Warburger
  Str.\ 100, D-33098 Paderborn, Germany}

\date{\today}

\begin{abstract}
The magnitude of finite-size effects and Coulomb interactions
in quantum Monte Carlo simulations of van der Waals interactions
between weakly bonded benzene molecules are investigated. 
To that extent, two trial wave functions of the Slater-Jastrow 
and Backflow-Slater-Jastrow types are employed to calculate
the energy-volume equation of state. We assess the impact of the
backflow coordinate transformation on the non-local correlation energy.
 We found that the effect of finite-size errors in quantum Monte Carlo calculations on
energy differences is particularly large and may even be more important
than the employed trial wave function. 
Beside the cohesive energy, the singlet excitonic energy gap and the  
energy gap renormalization of crystalline benzene at different densities are computed. 
\end{abstract}

\maketitle

\section {Introduction}
In his original work of 1930 \cite{London}, Fritz London introduced the fragment-based 
approach to van der Waals interactions. As a consequence, the dispersion energy between two 
spherical objects $A$ and $B$ can be calculated using second-order 
perturbation theory for the Coulomb interaction. 
In its general form, the vdW energy ($E_{vdW})$ can be written as 
\begin{equation}
E_{vdW} = - \sum_{n=6,8,10,...} f_{n}({\bf R}_{AB}, {\bf r}_{c,AB}) \frac{C_n}{{\bf R}^{n}_{AB}}, 
\end{equation}
where $f_{n}({\bf R}_{AB},{\bf r}_{c,AB})$ is a damping function, which depends on 
a cutoff distance ${\bf r}_{c,AB}$, whereas ${\bf R}_{AB}$ is the distance between
two fragments. The damping function attenuates the vdW energy for small values of
${\bf R}_{AB}$, where the electron clouds around fragments overlap\cite{Koide,Tang}. 
The dipolar term $C_{6}/{{\bf R}^{n}_{AB}}$ of the London expression has been widely used
for computing the vdW interactions within approximate first-principles electronic 
structure techniques, including Hartree-Fock (HF) \cite{Hepburn,Ahlrichs} and 
density functional theory (DFT) based methods \cite{Wu,Zimmerli,Johnson,Grimme,Tkat1,Woods}.
Although, this description of vdW energy is particularly accurate in the far-field 
for which the overlap between orbitals of the fragments is small, it is nevertheless very practical 
to improve the accuracy of generalized gradient approximation (GGA) exchange and correlation (XC) functionals.
In fact, it has been demonstrated that the inclusion of long-range dispersion effects systematically improves the 
description of non-local intermolecular interactions\cite{Grimme}.

The so-called vdW-DF approach has been extensively used to correct conventional local and semi-local
GGA XC functionals\cite{Dion,Cooper,Lee}. Therein, the XC energy is generally be expressed as 
$E_{XC} = E_{X}^{GGA}+E_{C}^{LDA}+E_{C}^{non-local}$, 
where $E_{X}^{GGA}$ is the exchange term from a given GGA XC functional. 
The local correlation energy $E_{C}^{LDA}$, however, is identical for most 
vdW functionals. By definition, the non-local part of the correlation energy
$E_{C}^{non-local}$ does not suffer from the Coulomb self-energy of each electron. 
Hence, XC self-interaction errors in vdW-DF schemes are mostly related to the corresponding exchange part. 
Similar to semi-local GGA XC functionals, the results of vdW-DF approaches depend on the employed $E_{X}^{GGA}$\cite{JCP16,PCCP}.
There are many non-covalent systems, where the accuracy of DFT falls short
of requirements. In particular, if the problem is to distinguish between molecular
crystal phases and competing low-energy polymorphs. As a simple molecular 
system, the energy differences between crystalline benzene and its polymorphs 
under pressure are less than a few kJ/mol.  It has been demonstrated that the use
of {\it ab initio} many-electron wave function methods, is essential to tackle this
 problem \cite{yang}. 

Interestingly, the similarity between Schr\"{o}dinger's equation 
in imaginary-time and the diffusion equation suggest to employ a 
stochastic diffusion-based process for solving the many-body Schr\"{o}dinger 
equation\cite{DMC,Ceperly1,Ceperly2}. In fact, quantum Monte Carlo 
(QMC) \cite{Matthew,MitasRev,Luchow}, which is family of stochastic methods for solving 
the Schr\"{o}dinger equation, is becoming an effective approach for 
investigating vdW interactions\cite{JCP15,JCP16,Dubecky}. In particular, previous studies have
shown that diffusion Monte Carlo (DMC) can provide accurate
energies for atoms\cite{Marchi9}, molecules\cite{QC} and crystals
\cite{Marchi11,JCP143,PRL,PRB17,JCP17,NJP} with non-covalent interactions
\cite{PCCP,Rezac,Zen18}. 

In order to mimic an extended system, QMC simulations of crystals are performed
using finite simulation cells subject to periodic boundary
conditions. Yet, practical and computational constraints restrict the
maximum size of the simulation cell and so introduce finite-size (FS)
errors, which can be rather large. The FS effects are larger in QMC than 
in mean-field methods because electrons are explicitly represented. 
Quantifying and minimizing these errors is an
essential part of all QMC simulations of extended systems,
particularly when high accuracy is required. 

In this work, we employ variational Monte Carlo (VMC)\cite{VMC}, as well as DMC\cite{DMC}, to 
study the Coulomb interaction between benzene rings in solid form with 
$Pbca$ symmetry. Crystalline benzene, due to its aromatic vdW interactions, 
is a model structure for studying non-covalent interactions in solids. 
Specifically, we calculate the vdW energy between four benzene molecules in a 
periodic simulation cell. We use the fragment-based approach, where the only degree of 
freedom is the distance between the center of mass of the benzene molecules. 

The energy gap of crystalline benzene has attracted a considerable amount of interest 
because of its importance in fundamental and applied science\cite{Wright}.
The energy gap $E_g$ is defined as the difference between the ionization energy and 
electron affinity. The Coulomb interactions between molecules, 
which are packed in a crystalline phase, reduces the fundamental gap as compared to the gas phase. 
This renormalized energy gap effect is crucial in organic electronics, especially in charge transport. 
The gas-phase energy gap is typically several eV, which is rather large for practical applications.  
However, DFT based methods are unable to quantify the fundamental 
energy gap of solid benzene and to distinguish the gas-phase gap from that of the crystallized 
structure\cite{Droghetti, Sivan}. Yet, the energy band gap of a molecular crystal, 
as determined by the GW approximation to many-body perturbation theory, is in agreement 
with experimental measurements\cite{Sharifzadeh,Neaton}. 

We employ the DMC approach to accurately calculate the excitonic energy gap of
crystalline benzene. Although the  DMC method was initially developed to 
study ground-state properties only, it can also be applied to determine 
excited states spectra in atoms, molecules, and crystals\cite{PRB17, Mitas94, Towler00}. 
The excitonic energy gap is smaller than the quasi-particle
band gap, which is usually determined by the GW approximation. The reason is the 
attraction between the excited electron in the conduction and the 
introduced hole in the valence band. The exciton binding energy is defined as 
the difference between the quasi-particle and the excitonic energy gaps. When an 
electron is added to (removed from) a finite simulation cell with periodic 
boundary conditions, a periodic lattice of quasi-particles (quasi-holes) is 
created. The energy of this lattice of quasi-particles is similar to 
the Madelung constant of the simulation cell lattice and introduces a large 
FS error in the electron affinity and ionisation potential.
The FS error in the quasi-particle energy gap is much larger than the excitonic 
gap for which the number of electrons is fixed. When an electron is added to 
(removed from) a finite and periodic simulation cell in which the interaction 
between particles is controlled by the Ewald potential, a neutralising 
charge density background is applied which vanishes in the infinite system 
size limit. Hence, the quasi-particle band gap can be physically meaningful in 
the infinite system size limit.

The remaining of this manuscript is organized as follows. Section~\ref{CD} describes 
the details of our VMC and DMC calculations. The corresponding results are discussed 
in section~\ref{RD}, which is followed by our conclusions in section~\ref{con}. 

\section {Computational Details}\label{CD}
The DMC method is a stochastic technique for
calculating the zero temperature total electronic energy of a many-electron system \cite{DMC}. Even though, 
DMC has been described in previous review articles\cite{Matthew,MitasRev,Luchow}, we will nevertheless start 
with a brief explanation of the general scheme since there are some technical aspects in this work we feel are rather important. 

More precisely, the DMC method solves the imaginary-time Schr\"{o}dinger equation 
\begin{equation}
  \frac{\partial \Psi(\mathbf{R}, \tau )}{\partial\tau} = \frac{1}{2}
  \sum^{N_e}_{i=1} \nabla_{\mathbf{r}_i}^2 \Psi(\mathbf{R}, \tau )
  - (V(\mathbf{R})-E_{T})\Psi(\mathbf{R},\tau),
\end{equation}
where $\mathbf{R} = (\mathbf{r}_1,\mathbf{r}_2,\ldots,\mathbf{r}_{N_e})$ 
is a $3N_e$-dimensional vector representing the positions of all $N_e$ 
electrons in the simulation cell, $\tau$ is the imaginary-time, $V(\mathbf{R})$ is the potential energy
including electron-electron interactions and $E_T$ is a constant energy
offset. Throughout, Hartree atomic units are assumed, i.e. the numerical values of
$\hbar$, $e$, $m_e$ and $4\pi\varepsilon_0$ are all identical to 1. As already alluded to above, the
imaginary-time Schr\"{o}dinger equation is similar to a $3N_e$-dimensional
diffusion equation with diffusion constant $D = 1/2$. The potential
energy term causes the diffusers to ``branch'' (multiply or die out) at
a position dependent rate. The wave function $\Psi(\mathbf{R},\tau)$ is
the number density of diffusers, which are normally known as walkers or
configurations and are points in the $3N_e$-dimensional configuration
space, not individual electrons. The DMC method employs this 
physical interpretation to simulate the imaginary-time evolution of the
wave function using a finite population of diffusing and branching walkers.

By solving the imaginary-time Schr\"{o}dinger equation, the electronic ground-state 
is projected out as $\tau \rightarrow \infty$. If the
initial wave function is expanded as a linear combination of energy
eigenfunctions $\Psi(\tau=0) = \sum_{i} c_i \Psi_i$, the solution of
the imaginary-time Schr\"{o}dinger equation $\partial\Psi/\partial\tau =
-(\skew3\hat{H}-E_T)\Psi$ is
\begin{equation}
  \Psi(\tau) = \sum_i c_i e^{-(E_i - E_T)\tau} \Psi_i.
\end{equation}
Thus, as long as $c_0 \neq 0$, the wave function $\Psi(\tau)$ becomes
proportional to $\Psi_0$ as $\tau \rightarrow \infty$. By gradually
adjusting $E_T$ to maintain the normalization of the solution in the
large $\tau$ limit, we can find the ground-state energy $E_0$.

Nevertheless, a fundamental difficulty with this approach is that the wave function
$\Psi(\mathbf{R},\tau)$, which is not necessarily positive, is
interpreted as a walker density that must be positive by its very definition. The 
naive application of the DMC algorithm to a many-electron system yields
a totally symmetric many-boson ground-state of no physical interest. The
so-called fixed-node approximation requires a trial 
wave function $\Psi_T(\mathbf{R})$, which imposes a fixed nodal 
constraint and hence prevents walker moves that cause
$\Psi_T$ to change sign. As long as $\Psi_T$ is properly antisymmetric,
this is sufficient to ensure that a fermionic solution is obtained. It 
can be shown that the energies calculated within
the fixed-node approximation are variational\cite{Matthew}: the result is greater than
or equal to the many-fermion ground-state energy and tends to the exact
energy as the $(3N_e-1)$-dimensional nodal surface, on which $\Psi_T=0$,
approaches the ground-state nodal surface. Even though, assuming the fixed-node approximation is
essential for DMC simulations of large systems, it is the only
fundamental limitation of the method. Other approximations, such as the
use of a finite time-step or the representation of ions by
pseudopotentials, can be made negligible or fully avoided given sufficient
computer time. 

The diffusion and branching process as described above is unstable in
practice since the potential energy $V(\mathbf{R})$ diverges whenever
electrons approaches nuclei or each other, leading to an uncontrollable
branching. This problem, however, can be overcome using an importance-sampling
technique. To that extent, the imaginary-time Schr\"{o}dinger equation is rewritten 
in terms of the quantity $f(\mathbf{R},\tau) =
\Psi_T(\mathbf{R})\Psi(\mathbf{R},\tau)$ to obtain
\begin{eqnarray}\nonumber
  \frac{\partial f(\mathbf{R},\tau)}{\partial t} &=&
  \frac{1}{2}\nabla_{\mathbf{R}}^2 f(\mathbf{R},\tau) -
  \bm{\nabla}_{\mathbf{R}}\cdot \left [ \mathbf{v}(\mathbf{R})
    f(\mathbf{R},\tau) \right ] -\\
 && [E_L(\mathbf{R}) - E_T]f(\mathbf{R},\tau),
\end{eqnarray}
where $\bm{\nabla}_{\mathbf{R}} = (\bm{\nabla}_{\mathbf{r}_1},
\bm{\nabla}_{\mathbf{r}_2},\ldots, \bm{\nabla}_{\mathbf{r}_{N_e}})$ is the
$3N_e$-dimensional gradient operator, $\nabla_{\mathbf{R}}^2 =
\bm{\nabla}_{\mathbf{R}}\cdot\bm{\nabla}_{\mathbf{R}}$ is the
corresponding Laplacian, $\mathbf{v}(\mathbf{R}) =
\bm{\nabla}_{\mathbf{R}}\ln|\Psi_T(\mathbf{R})|$ is the $3N_e$-dimensional
drift-velocity vector and $E_L(\mathbf{R}) = (1/\Psi_T(\mathbf{R}))
\hat{H}\Psi_T(\mathbf{R})$ is the local energy. The importance-sampled
imaginary-time Schr\"{o}dinger equation corresponds to a diffusion process
similar to that discussed above, except that the walkers now drift with
velocity $\mathbf{v}(\mathbf{R})$, as well as diffusing and
branching. The branching rate is determined by the shifted local energy
$E_L(\mathbf{R})-E_T$ instead of the shifted potential energy
$V(\mathbf{R})-E_T$. If the trial function is a good approximation to
the ground-state, the local energy is a smooth function of $\mathbf{R}$
and the numerical difficulties caused by divergences in $V(\mathbf{R})$
are bypassed. The fixed-node approximation is imposed by rejecting
walker moves that change the sign of $\Psi_T(\mathbf{R})$.

In this work, the \textsc{casino} code\cite{casino} was used to perform 
DMC simulations with a trial wave function of the Slater-Jastrow (SJ) form 
\begin{equation}
  \Psi_{\rm SJ}({\bf R}) = \exp[J({\bf R})] \det[\psi_{n}({\bf r}_i^{\uparrow})] 
  \det[\psi_{n}({\bf r}_j^{\downarrow})],
\label{WF}
\end{equation}
where ${\bf R}$ is a $3N$-dimensional vector containing the positions of all $N$ electrons,
${\bf r}_i^{\uparrow}$ the position of the $i$'th spin-up electron, ${\bf r}_j^{\downarrow}$
the position of the $j$'th spin-down electron, $\exp[J({\bf R})]$ the Jastrow correlation factor, 
while $\det[\psi_{n}({\bf r}_i^{\uparrow})]$ and $\det[\psi_{n}({\bf r}_j^{\downarrow})]$ 
are Slater determinants made of spin-up and spin-down one-electron wave functions. 
These orbitals were obtained from PBE-DFT calculations performed with the CASTEP
plane-wave code\cite{castep}, in conjunction with Trail-Needs Dirac-Fock 
pseudopotentials\cite{TN1,TN2}. For the purpose to approach the complete basis set
limit\cite{sam}, a large energy cut-off of 4000~eV have been chosen. 
The resulting plane-wave orbitals were subsequently transformed into a
localized blip polynomial basis\cite{blip}. Our DMC results were obtained 
using a real $\Gamma$-point wave function. 

The Jastrow correlation factor within Eq.~\ref{WF} is a positive, symmetric, explicit function of
interparticle distances in the form of:
\begin{eqnarray}\nonumber
J({\bf r}_i, {\bf r}_I)&=&
\sum_{I=1}^{M}  \sum_{i=1}^{N}                  u_1({\bf r}_{iI})+
\sum_{i=1}^{N-1}\sum_{j=i+1}^{N}                 u_2({\bf r}_{ij})+\\
&& \sum_{I=1}^{M}  \sum_{i=1}^{N-1}\sum_{j=i+1}^{N} u_3({\bf r}_{iI},{\bf r}_{jI},{\bf r}_{ij}), 
\label{Jas}
\end{eqnarray}
where $N$, $M$, ${\bf r}_i$ and ${\bf r}_I$ are the number of electrons, number of ions,
the position of electron $i$ and the position of nucleus $I$, 
whereas ${\bf r}_{ij}={\bf r}_i - {\bf r}_j$ and ${\bf r}_{iI}={\bf r}_i - {\bf r}_I$. 
The polynomial one-body 
electron-nucleus (1b), two-body electron-electron (2b) and three-body 
electron-electron-nucleus (3b) terms are denoted as $u_1(r_{iI})$,$u_2(r_{ij})$ and $u_3(r_{iI},r_{jI},r_{ij})$, respectively.

We also studied the contribution of nondynamical correlation by including the inhomogenous backflow (BF)
coordinate transformation into the SJ wave function\cite{Lopez}. Our BF transformation includes
electron-electron correlation factor as well as electron-proton terms and is given by
\begin{equation}
\mathbf{X}_i(\{\mathbf{r}_j\})=\mathbf{r}_i+\bm{\xi}^{(e-e)}_i(\{\mathbf{r}_j\})
+\bm{\xi}^{(e-P)}_i(\{\mathbf{r}_j\}),
\end{equation}
where $\mathbf{X}_i(\{\mathbf{r}_j\})$ is the transformed coordinate of electron $i$, 
which depends on the full configuration of the system $\{\mathbf{r}_j\}$. 
The vector functions $\bm{\xi}^{(e-e)}_i(\{\mathbf{r}_j\})$ and 
$\bm{\xi}^{(e-P)}_i(\{\mathbf{r}_j\})$ are the electron-electron and electron-proton
backflow displacements of electron $i$. They are parameterized as
\begin{equation}
\bm{\xi}^{(e-e)}_i(\{\mathbf{r}_j\})=\sum_{j\neq i}^{N_{e}} \alpha_{ij}(r_{ij}) \mathbf{r}_{ij}
\end{equation}
and
\begin{equation}
\bm{\xi}^{(e-P)}_i(\{\mathbf{r}_j\})=\sum_{I}^{N_{P}} \beta_{iI}(r_{iI}) \mathbf{r}_{iI}, 
\end{equation}
where $\alpha_{ij}(r_{ij})$ and $\beta_{iI}(r_{iI})$ are polynomial functions of  
electron-electron and electron-proton distances, respectively, and contains variational parameters.
In this way, the resulting backflow SJ (BSJ) is able to adapt the nodal surface in order to
recover the static correlation energy that is characteristic for multi-reference systems\cite{EPL}.
All adjustable parameters in the Jastrow correlation factor and BF coordinate transformation are optimized 
by minimizing the variance, as well as the variational energy at the VMC 
level\cite{varmin1,varmin2}. The Kato 
cusp conditions are enforced so that the local energy is finite when 
two electrons or an electron and a nucleus are coincident\cite{Kato, PackBrown}. Specifically, 
the electron-electron cusp condition are imposed on the parameters of 
the Jastrow correlation factor and the electron-nucleus ones on the 
orbitals within the Slater determinant. Since the backflow coordinate transformation can modify the cusp conditions, we have constrained the 
backflow parameters so that they are not\cite{Lopez}.

The excitonic energy band gap is determined by promoting 
an electron from a valence band state into a conduction band 
orbital at the $\Gamma$ point. The singlet excited state was defined by promoting an 
electron without flipping its spin:
\begin{equation}
\Delta_{exc} = E_1 - E_0,
\end{equation}
where $E_0$ and $E_1$ are the DMC energies of the ground and excitonic states, respectively.
The excitonic energy equals to the vertical optical absorption gap.  

The FS errors are categorized into one- (independent particle)
and many-body terms. The one-body term includes the non-interacting kinetic, potential, 
as well as Hartree energies. The one-body FS errors are much more important in metallic 
systems due to shell-filling effects\cite{Neil08}. The single-particle errors in metallic 
systems are eliminated using canonical and grand-canonical twist averaging boundary
 conditions\cite{Lin01,Holzmann}. Many-body 
FS errors are due to the exchange and correlation effects within the Coulomb and kinetic 
energies and can not be removed by twist averaging. Crystalline benzene is a wide band 
gap insulator, which is why many-body FS effects are the main source of errors in our 
DMC calculations. Hence, we investigate the influence of these FS errors on the vdW interaction 
between aromatic benzene rings. There are different approaches 
to reduce or cancel many-body errors. The most widely used and perhaps oldest approximation 
is extrapolating to the infinite system size limit, which however is computationally rather expensive 
for the large simulation cells that are studied here. 
We therefore analyse two alternative methods to correct for many-body 
FS errors in our DMC simulations. Specifically, we apply the structure-factor-based approaches proposed 
by Chiesa \emph{et al.}\cite{Chiesa}, which allows to estimate the effect of 
many-body FS errors in the potential and kinetic energies based on the random phase 
approximation. The main assumption in this approach is that the low-\emph{k} behaviour 
of the structure factor is independent of the shape of the simulation cell. Here, we first 
apply the standard Ewald form of the periodic Coulomb interaction and Chiesa's FS corrections 
for both, the kinetic ($\Delta$KE) and potential ($\Delta$PE) energies. Second, we employ the 
Model Periodic Coulomb (MPC) interactions\cite{Fraser,Williamson} to deal with the Coulomb 
errors. Our DMC results obtained with both of these approaches are expected to be similar. 
\section {Results and discussion}\label{RD}
\begin{figure}
\begin{tabular}{c c}
\includegraphics[width=0.245\textwidth]{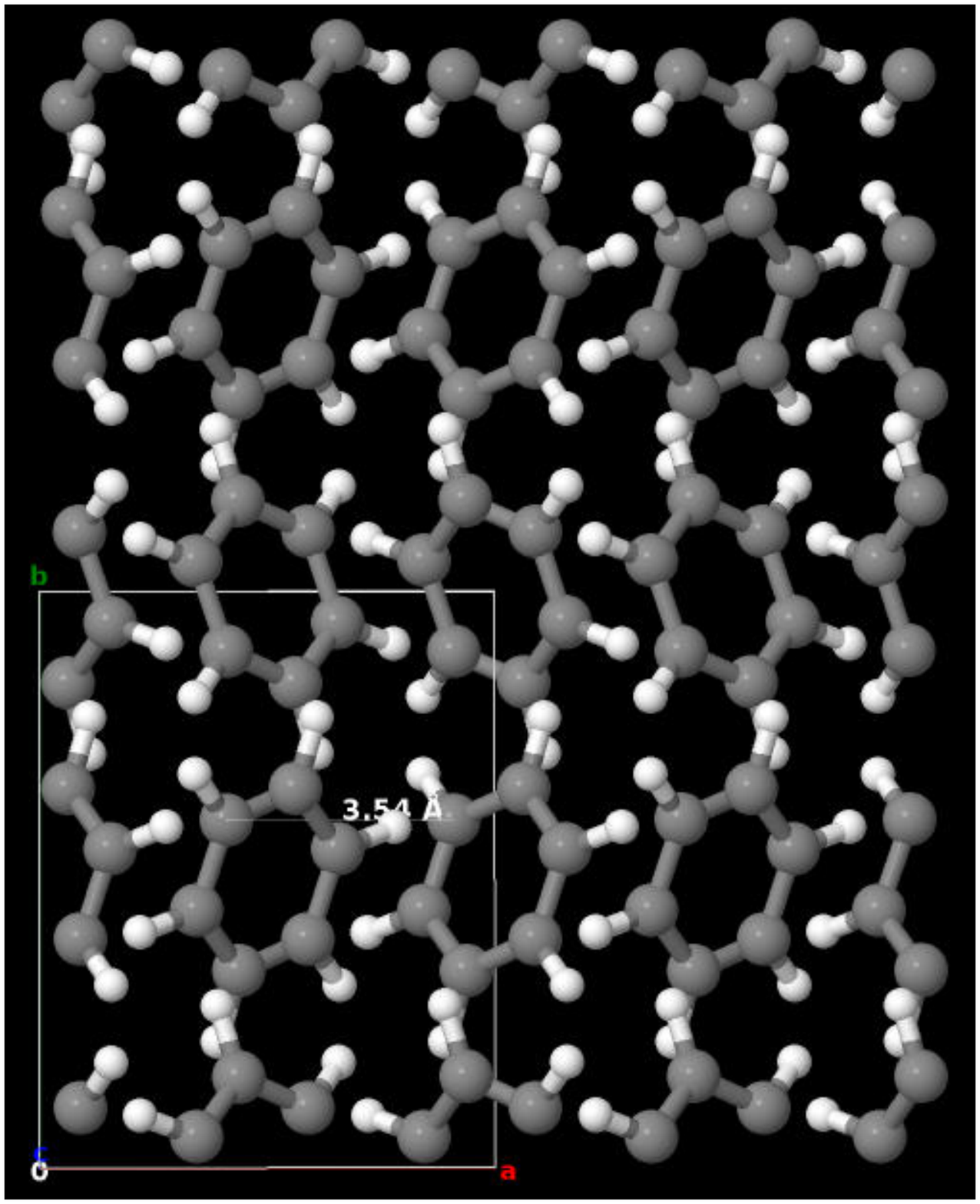}&
\includegraphics[width=0.25\textwidth]{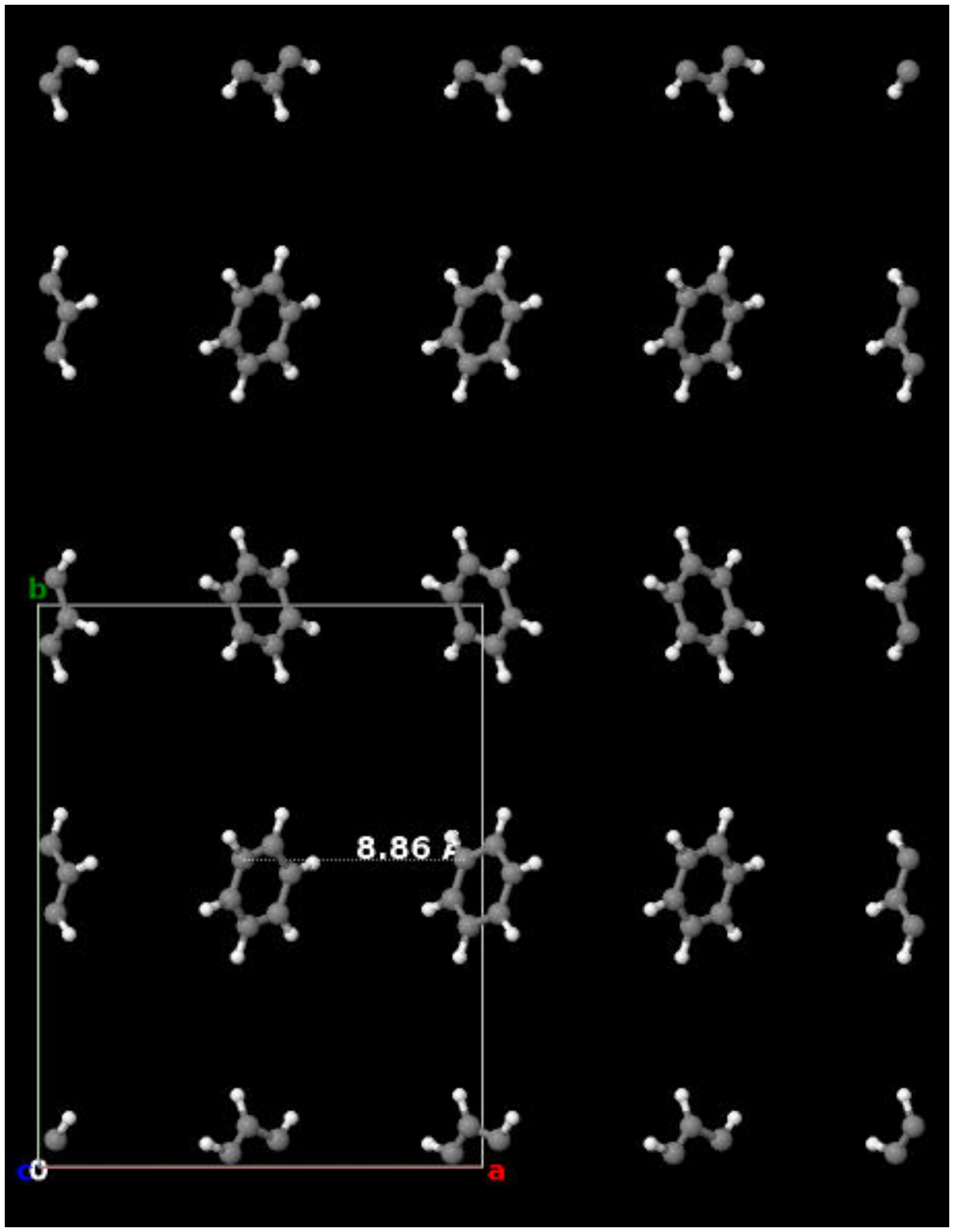}\\
\includegraphics[width=0.25\textwidth]{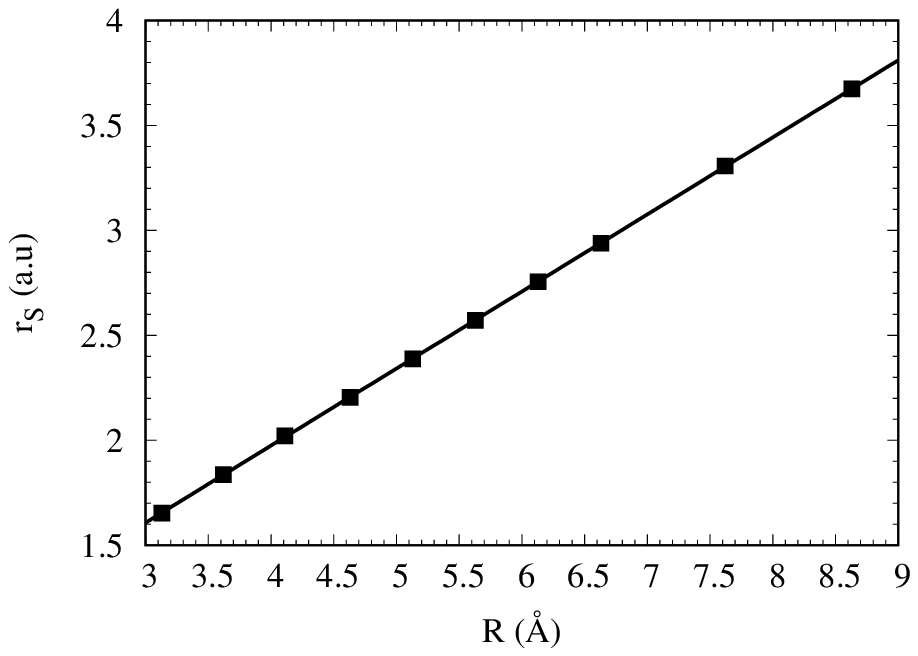} &
\includegraphics[width=0.25\textwidth]{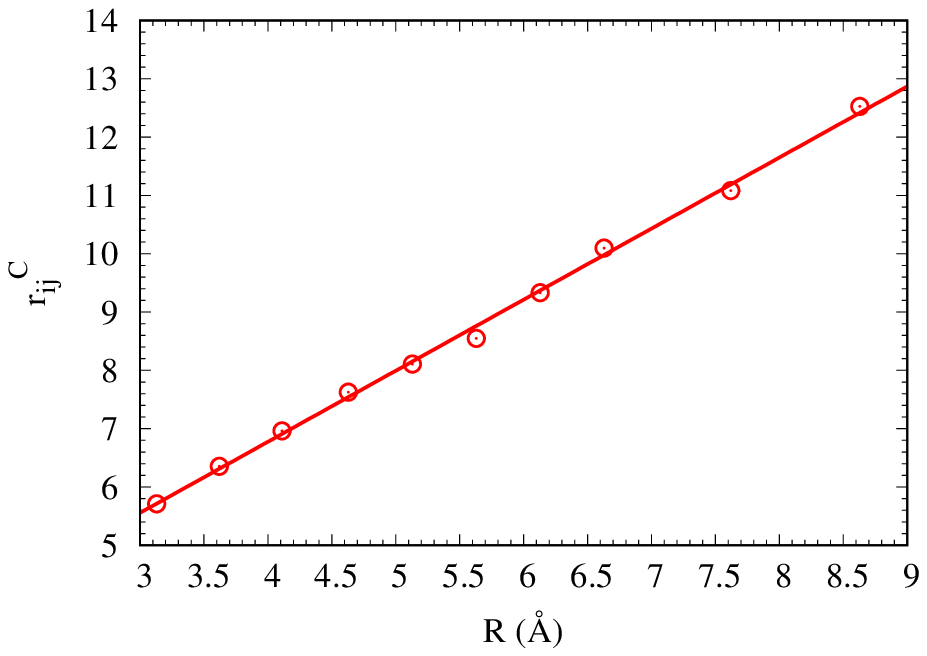} \\
\end{tabular}
\caption{\label{cell} (Color online) (Top) The smallest and largest simulation cells
with volumes $V_s = 336.4334 \AA^3$ and $V_l = 3691.9989 \AA^3$, respectively, and which are 
used to calculate the vdW interactions between the benzene molecules. (Bottom) Wigner-Seitz radius $r_S$ and 
cutoff length $r_{ij}^C$ as linear functions of R. } 
\end{figure}

The orthorhombic simulation cell with $Pbca$ symmetry contains four
benzene molecules and 120 electrons. More precisely, we are considering 
10 different simulation cell sizes, whose volumes are between $V_s = 336.4334 \AA^3$ and 
$V_l = 3691.9989 \AA^3$, respectively. The volume of the simulation 
cells was varied in such way that the distance $R$ between the center  
of mass of the benzene molecules is a linear function of $r_S$ (electronic 
Wigner-Seitz radius), as it is shown in Fig.~\ref{cell}. The relative 
configurations of the benzene molecules and their geometries were fixed. 

In total, there are 151 and 225 variational parameters in our SJ and BSJ wave functions, respectively.
We first optimized the expansion coefficients of the Jastrow correlation factor of the SJ wave function. The 
optimized coefficients were subsequently used to also optimize of cutoff length in the Jastrow term.
The optimized Jastrow term was reused to generate the BSJ wave function. In our BSJ 
wave function optimization, we first optimized the variational parameters of the BF coordinate transformation, while 
the parameters in the Jastor correlation factor were kept fixed, before we reoptimized all the variational parameters together.
We found that the described optimization procedure produces an accurate wave function for DMC calculations.  
The two-body $u_2({\bf r}_{ij})$ term consists of a power expansion in $r_{ij}$ and goes 
to zero at the cutoff length $r_{ij}^C$. We found that our optimized values for $r_{ij}^C$ represent 
essentially a linear function of R. 
\begin{figure}
\includegraphics[width=0.5\textwidth]{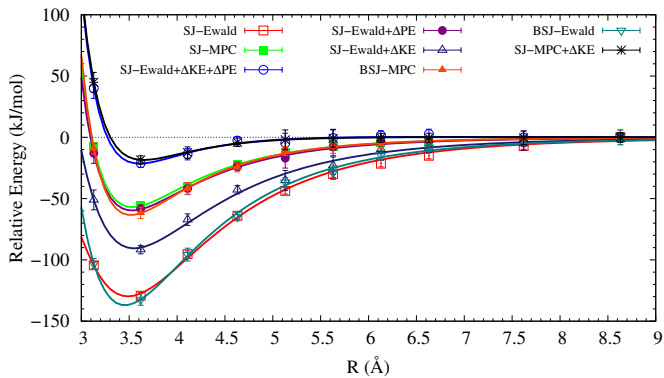}
\caption{\label{DMCvsd} (Color online) Relative DMC energy 
as calculated using the Ewald, and MPC potentials and SJ, and BSJ 
wave functions. The data points 
are fitted to $\frac{C_6}{R^6}+\frac{C_8}{R^8}+\frac{C_{10}}{R^{10}}$, 
where $C_n$ with $n=6, 8, 10, ...$ are fitting parameters and $R$ is the 
smallest distance between the center of mass of the benzene molecules.
The reference line is the DMC energy at the largest separation.} 
\end{figure}

The resulting non-local vdW energy curves between four benzene molecules, which are calculated using the
Ewald and MPC interactions, are illustrated in Fig.~\ref{DMCvsd}.
The DMC data points are reported in the Supplementary Information\cite{supp}. 
An advantage of the MPC approach is that it reduces FS 
errors arising from the use of the Ewald interaction. As an alternative to the 
MPC scheme, many-body contributions to the FS errors can also be minimized 
using the FS correction to the exchange-correlation ($\Delta$PE) and kinetic ($\Delta$KE) energies\cite{Chiesa}.
The total energy of the system at the largest separation is considered as the zero reference. 
We would like to emphasize that in our approach, the size-consistency problem within binding energy 
calculations\cite{Zen}, is avoided. The results of the BSJ wave function with the Ewald and
MPC interactions are also shown in Fig.~\ref{DMCvsd}. As can be seen, 
employing the MPC interaction corrects the Coulomb FS errors, but not the many-body FS error of the kinetic energy. 
Also, the binding energy curve, as calculated by SJ-Ewald$+\Delta$PE scheme, 
agrees well with the corresponding SJ-MPC results. Even though, the magnitude of $\Delta$KE is smaller than that of $\Delta$PE, 
combining both FS correction techniques entails the largest contribution. 
Interestingly, the results using the SJ and BSJ wave functions are remarkably similar with each other. 
The Ewlad+$\Delta$KE+$\Delta$PE results are in excellent agreement with the DMC energies which are 
calculated using MPC+$\Delta$KE. Note that the higher-order kinetic energy corrections 
defined according to Eq. (55) in Ref. \onlinecite{Neil08} are included in $\Delta$KE. 
The DMC energy curves were all fitted to $\frac{C_6}{R^6}+\frac{C_8}{R^8}+\frac{C_{10}}{R^{10}}$, 
where $C_n$ with $n=6, 8, 10, ...$ are fitting parameters and $R$ is the 
smallest distance between the center of mass of the benzene molecules. 

\begin{table}
\centering
\begin{tabular}{c c c}
\hline\hline
Approach                                              &    Energy  & Reference   \\ \hline
DMC SJ Ewald                                       &  -125.4$\pm$9.6   & this work \\ 
DMC SJ MPC                                         &  -57.9$\pm$9.6    & this work \\
DMC SJ Ewald+$\Delta$PE                    &  -57.9$\pm$9.6    & this work \\
DMC SJ Ewald+$\Delta$KE                    &  -86.8$\pm$9.6    & this work \\
DMC SJ Ewald+$\Delta$PE+$\Delta$KE &  -29.0$\pm$9.6    & this work \\
DMC SJ MPC+$\Delta$KE                      &  -29.0$\pm$9.6    & this work \\
DMC BSJ Ewald                                    &  -135.1$\pm$9.6     & this work \\
DMC BSJ MPC                                      &  -67.5$\pm$9.6     & this work \\
DMC SJ MPC                                         &  -52.1$\pm$0.4     &\onlinecite{Zen18} \\
Estimated Experiment at 0K                  &   -55.3$\pm$2.2    &\onlinecite{yang}   \\
Quantum Chemistry (CCSD(T,Q))        &  -55.9$\pm$0.76$\pm$0.1&\onlinecite{yang}  \\
B3LYP-D Grimme/6-31G(d,p)               & -48.2$\pm$20.1      &\onlinecite{Civalleri}  \\
B3LYP-D Grimme/TZP                          & -46.5$\pm$1.9       &\onlinecite{Civalleri}   \\
B3LYP/6-31G(d,p)                                & -5.8$\pm$8.9        & \onlinecite{Civalleri} \\
CCSD(T)/CBS                                       & -56.4                     &\onlinecite{Ringer}\\
Typical experimental values                  & -43 to -47             &\onlinecite{Chickos}\\
DFT LDA                                               & -57.00                     &  \onlinecite{DLu} \\
EXX/RPA (LDA)                                     & -44.00                  &   \onlinecite{DLu} \\
DFT PBE                                                & -9.60                   &  \onlinecite{DLu}  \\
EXX/RPA (PBE)                                      & -47.00                 &   \onlinecite{DLu} \\
DFT/LDA+B                                            & -34                      &  \onlinecite{Meijer}   \\
DFT/LDA+B+LYP                                     & -65                      &  \onlinecite{Meijer}     \\
\hline\hline
\end{tabular}
\caption{\label{Ecoh} The cohesive energy of crystalline benzene as obtained by 
	different DMC approaches at 0~K. All energies are in kJ/mol. The quantum mechanical zero-point energy contribution,  
	which amounts to 2.8 kJ/mol \cite{Nakamura}, is not included in our results. }
\end{table}
We computed the binding energy between the aromatic rings, which indicates
the strength of the vdW forces holding the benzene molecules together. 
The cohesive energy ($E_{coh}$) is defined as $E_{coh}= E_{DMC}^{R_0}-E_{DMC}^{R_\infty}$,
where $E_{DMC}^{R_0}$ and $E_{DMC}^{R_\infty}$ are the DMC energies of the system 
at the equilibrium distance $R_0$ and infinite separation $R_\infty$, respectively.
For the later, we assume that $R_\infty= 8.64 \AA$.
Our cohesive energies, as determined using different DMC schemes, are listed 
in table~\ref{Ecoh}. We find that using the bare Ewald interaction, the cohesive energy 
is severely overestimated due to presence of significant FS errors. 
However, correcting for the latter by including $\Delta$PE and $\Delta$KE, 
the cohesive energy is reduced by as much as 96~kJ/mol. 
Moreover, comparing the contributions of $\Delta$PE and $\Delta$KE immediately
suggests that the former is more effective and outperforms the 
$\Delta$KE FS correction by 29~kJ/mol. In fact, the $\Delta$PE FS 
correction is equally effective than employing the MPC approach. 
Although including the BF coordinate transformation generally improves the ground-state total energy, 
the cohesive energies, as obtained by the SJ Ewald and BSJ Ewald schemes, differs by just 9.5(1)~kJ/mol, 
which is also the case when adding the MPC interaction. 
This is to say that when calculating energy differences using QMC, 
the impact of FS errors is particularly important and may even be 
more important than the particular trial wave function. 
Recently, the DMC method has been applied to calculate the EOS of 
	molecular crystals\cite{Zen18}. In particular, the authors calculated the lattice energy of crystalline 
	benzene using: $E_{latt} = E_{crys} - E_{gas}$, which is listed in Table \ref{Ecoh}.
	Our results that are obtained using the MPC potential agree well with those recently reported
	ones. 
\begin{figure}
\includegraphics[width=0.45\textwidth]{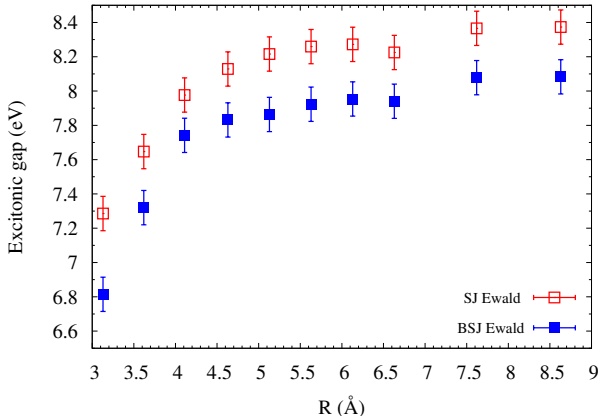}
\caption{\label{gap} (Color online) Singlet DMC excitonic gap, which 
are determined using the SJ and BSJ wave functions in conjunction 
with the bare Ewald potential, as a function of $R$.}
\end{figure}

The singlet DMC excitonic gaps, which were obtained at 
different densities using the SJ and BSJ wave functions, are shown in Fig.~\ref{gap}. 
The highest density (smallest $R$) corresponds to crystalline benzene at 8~GPa pressure,
while the lowest density (largest $R$) mimics the gas 
phase. On the one hand, the energy gaps computed by means of HF theory are typically too large 
due to the absence of electron-electron correlation. On the other hand, the band gaps 
calculated by conventional DFT methods are generally too small. In DMC simulations, however, 
multiplying the Slater determinant made up of HF or DFT orbitals, by a Jastrow correlation 
factor permits to retrieve a high amount of the correlation energy and results in energy gaps much closer to 
experiment. But, applying the Jastrow correlation factor 
does not alter the nodal surface of the trial wave function, which is determined by the Slater determinant.
Nevertheless, introducing a BF coordinate transformation to the SJ wave function affects the nodal surface 
and improves the ability of DMC to recover nondynamical correlation energy. As we found,
adding a BF coordinate transformation lowers the excitonic energy gap. When calculated by 
the GW approximation, the corresponding HOMO-LUMO gaps of benzene in the gas and crystal phases 
are 10.51 eV and 7.91 eV, respectively\cite{Neaton}. By contrast, at the DFT-LDA level of theory, the electron 
addition and removal energies of benzene in the gas and crystal phases are 5.16 and 5.07 eV, respectively.
For comparison, the experimental ionization potential of a benzene molecule is 9.25 eV\cite{Curtiss}.
Using modified hybrid and constrained DFT calculations \cite{Sivan, Droghetti}, 
the fundamental gap renormalization was previously found to be about $\sim$2~eV.   
The SJ-DMC calculations predict that the energy gaps at optimized and largest R are
7.7(1) and 8.4(1)~eV, respectively. The energy gaps which are obtained using BSJ-DMC simulations are
7.4(1) and 8.1(1)~eV, respectively. We calculated the gap renormalization using the difference between 
energy gaps at optimized R and the largest R.
Hence DMC excitonic energy gap results yield a gap renormalization of 0.7(1)~eV. 

However, DMC simulations of excitations spectra in solids are rather challenging due to an $1/N$
effect: the change in the total energy induced by an 
one- or two-particle excitation is inversely proportional to the number of 
electrons in the simulation cell. Thus, generally, a relatively large simulation cell is 
essential for a high-precision description of the infinite crystal. Yet, our dispersion 
energy curves demonstrates the importance to explicitly account for FS errors, in which 
case the interactions between benzene molecules is rather accurately described. 

\section{Conclusions}\label{con}
In this work, the non-local vdW interactions between benzene molecules 
was studied by means of DMC simulations using the SJ and BSJ wave functions. 
We found that, when calculating energy differences, the results are much more affected 
by FS errors than generally appreciated. 
In the case of the cohesive energy, FS errors can be as large as 96~kJ/mol,
which is much more pronounced than the impact of the BF coordinate transformation 
to include nondynamical correlation effects. In addition, we also calculated the 
singlet excitonic energy gap for benzene in the gas and solid phases. 
At variance to the cohesive energy, the inclusion of BF into the trial wave function
entails a reduction of the excitonic band gap. Eventually, we also obtained a high 
accuracy estimation of the benzene gap renormalization.  

\section{Acknowledgements}\label{ack}

The authors would like to thank the Gauss Center for Supercomputing (GCS)
for providing computing time through the John von Neumann Institute for
Computing (NIC) on the GCS share of the supercomputer JUQUEEN at the
J\"ulich Supercomputing Centre (JSC). Additional computing facilities
were provided through the DECI-13 PRACE project QMCBENZ15 and the 
Dutch national supercomputer Cartesius. This project has received funding
from the European Research Council (ERC) under the European Union's Horizon
2020 research and innovation programme (grant agreement No 716142).

\end{document}